\documentclass[12pt,preprint]{aastex}
\usepackage{graphicx}

\shorttitle{Plasma flows in active region moss}
\shortauthors{Durgesh Tripathi et al.}

\begin{document}

\title{Active Region Moss: Doppler Shifts from Hinode/EIS Observations}
\author{Durgesh Tripathi}
\affil{Inter-University Centre for Astronomy and Astrophysics, Pune University Campus, Pune 411007, India}

\author{Helen E. Mason}
\affil{Department of Applied Mathematics and Theoretical Physics, University of
Cambridge, Wilberforce Road, Cambridge CB3 0WA, UK}

\author{James A. Klimchuk}
\affil{NASA Goddard Space Flight Center, Greenbelt, MD 20771, USA}

\begin{abstract}Studying the Doppler shifts and the temperature dependence of Doppler shifts in moss regions
can help us understand the heating processes in the core of the
active regions. In this paper we have used an active region
observation recorded by the Extreme-ultraviolet Imaging Spectrometer
(EIS) onboard Hinode on 12-Dec-2007 to measure the Doppler shifts in
the moss regions. We have distinguished the moss regions from the
rest of the active region by defining a low density cut-off as
derived by \cite{tmdy_2010}. We have carried out a very careful
analysis of the EIS wavelength calibration based on the method
described in \cite{young_brendan_mason}. For spectral lines having
maximum sensitivity between $\log\,T = 5.85$ and $\log\,T = 6.25$ K,
we find that the velocity distribution peaks at around 0~km~s$^{-1}$
with an estimated error of 4$-$5~km~s$^{-1}$. The width of the
distribution decreases with temperature. The mean of the
distribution shows a blue shift which increases with increasing
temperature and the distribution also shows asymmetries towards
blue-shift. Comparing these results with observables predicted from
different coronal heating models, we find that these results are
consistent with both \textsl{steady} and \textsl{impulsive} heating
scenarios. However, the fact that there are a significant
number of pixels showing velocity amplitudes that exceed the
uncertainty of 5~km~s$^{-1}$ is suggestive of impulsive heating.
Clearly, further observational constraints are needed to distinguish
between these two heating scenarios.
\end{abstract}

\keywords{Sun: corona --- Sun: atmosphere --- Sun: transition region --- Sun: UV radiation}

\section{Introduction}

The problem of solar coronal heating has provided a major
challenge since the discovery of the hot
corona in 1940s. The solution remains elusive,
in spite of major advances in theoretical modelling and
observational capabilities in recent years. For an overview of our
current understanding of this problem, the reader is
referred to \cite{klimchuk_2006}.

Active regions are the brightest long-lived regions on the Sun and
present an ideal target of opportunity for studing the coronal
heating problem. Active regions are comprised of a diffuse component
\citep{vk2011} and a variety of observationally discrete structures,
including fan loops, warm loops, and hot core loops. The fan loops
(only the legs are visible) are large structures at the edges of
active regions, which are characterised by cool emission ($<$1MK),
the warm loops are clearly defined at around 1MK, for example in
Transition Region and Coronal Explorer \citep[TRACE]{trace},
173~{\AA} images \citep[e.g.][]{giulio_helen} and the hot loops are
seen in higher temperatures emission (3MK) for example in X-rays
(with Hinode/XRT and Yohkoh/SXT). The moss emission, also clearly
seen in TRACE~173~{\AA}, is believed to be the footpoints of the hot
core loops \citep[e.g.][and references therein]{tmdy_2010}.

One of the main concerns regarding the coronal heating problem is
whether all the different structures are heated via the same
mechanism or whether there are different processes at work for
different structures.

The observed properties of warm loops appear to be consistent with
models of \textsl{multi-stranded loops} heated by low-frequency
nanoflares (nanoflares with repetition time lags longer than a
cooling time on each strand) \citep[see e.g.,][and references
therein] {warren_2003, tripathi_2009, inaki_2009, klimchuk_2009}.
However, the results for the hot core loops have not converged.
There is compelling observational evidence for both steady heating
\citep[see e.g.,][]{warren_2008, brooks_warren, warren_2010,
tmdy_2010, wine_2011} as well as impulsive heating
\citep[][]{tmk_2010, tkm_2011,vk2012}. Since it is not easy to
observe the hot core loops directly, due to their inherent fuzzy
nature \citep{tripathi_2009, fuzzy_2010}, authors have focussed on
studying the characteristics and dynamics of the moss regions
\citep[see e.g.][and references therein]{tmdy_2010}. The moss
regions provide important constraints on the heating of the hot core
loops \citep[e.g.][]{wine_2008, warren_2010, tmdy_2010}. One of the
main arguments for steady heating is the minimal variability
observed in the brightness of the moss emission \citep{antiochos},
Doppler shifts, line widths \citep{brooks_warren} and electron
density \& temperature structure \citep{tmdy_2010}. However, we have
argued that this can be uniquely interpreted only if the plasma does
not have unresolved structures \citep{tmk_2010}.

One of the main hindrances to solving the coronal heating problem is
that we are still far from resolving the fundamental structures in
the solar corona \citep[e.g.][]{klimchuk_2006, warren_2008,
tripathi_2009, tmdy_2010, young_brendan_mason}. Therefore, the
observationally derived properties of coronal structures relate to
an ensemble of structures, rather than a fundamental structure
itself. This could have severe consequences for the interpretation of
the data. For example any small scale variations taking place in the fundamental structures could be completely
washed out/averaged in our present day observations. Therefore, we
have to rely on the diagnostic techniques which do not require the
fundamental structures to be resolved. There are two main diagnostic
methods for this purpose, the first being the emission measure (EM)
distribution and the second being the Doppler shift of the plasma in
the coronal structures.  A third method involving time-lag analysis
has recently been developed \citep{vk2012}.

In this work we have made use of the Doppler shift techniques and
studied the bulk flow of plasma in the moss regions, as advocated by
\cite{tmk_2010}. This is potentially a powerful diagnostic tool to
distinguish between steady heating and impulsive heating models.
Models advocating steady heating in symmetric loops predict no
Doppler shift because the loops are in a static equilibrium where
heating, radiation, and thermal conduction are perfectly balanced.
Asymmetries in the heating and/or cross sectional area will generate
steady flows \citep[e.g.,][]{mariska_boris}, but these flows will be
slow unless the asymmetries are extreme \citep{wine_2002, pk_2004}.
Furthermore, the flows are end-to-end and produce blue shifts in one
leg and red shifts in the other. Flows can also be generated in
symmetric loops if the heating is very highly concentrated at the
loop footpoints \citep[e.g.][]{klimchuk_2010,letal_2012}.  These
flows occur because of thermal nonequilibrium and are time-dependent
even though the heating is steady.

Impulsive heating produces a very different Doppler shift signature.
The standard picture envisions a collection of unresolved strands
that are heating quasi-randomly by nanoflares \citep[see
e.g.,][]{cargill_1994}.  An observed emission line then represents
the combined emission from many out-of-phase strands. Upflows are
generated by chromospheric evaporation, but these tend to be faint
because the evaporation phase is short lived and because the initial
densities are low. Downflows associated with the subsequent slow
cooling and draining of the evaporated plasma tend to be brighter
and therefore to dominate the emission. The composite line profile
is therefore predicted to be a red-shifted core with an enhanced
blue wing \citep{pk_2006}. The Doppler shift that
would be measured from a Gaussian fit will depend on the relative
strengths of the upflow and downflow emission.  These change
depending on the temperature of the emission line.

One goal of our study is to determine how Doppler shift
depends on temperature.  Constant mass flux at constant pressure
implies that $v \propto T$.  While we expect this relationship to
hold on each strand at each moment in time, the relationship between
Doppler shift and temperature for an unresolved collection of
strands is not straightforward.  Our preliminary modeling suggests
that coronal nanoflares produce red shifts which decrease with
temperature and eventually give way to blue shifts in hot lines
(Bradshaw and Klimchuk, in preparation).  The temperature of the transition
depends on the strength of the nanoflares.

Hansteen et al. (2010) have proposed a different impulsive heating
scenario. They suggest that nanoflares occurring in the upper
chromosphere cause material to expand downward and upward away from
the locally enhanced pressure.  This produces red shifts in
transition region emissions and blue shifts in coronal emissions.

\cite{brooks_warren} studied the Doppler shift and non-thermal width
and their temporal variation in moss regions using the
\ion{Fe}{12}~$\lambda$195 line observed by the EIS. They reported a
minimal red-shift of about 2-3~km~s$^{-1}$ and almost no variability
in Doppler shift and non-thermal width with time. These results led
them to conclude that the hot core loops are heated in steady
fashion. However, it is worthwhile mentioning here that the
reference wavelength they choose for Doppler measurements was
averaged over the complete raster. If the emission outside the core
has a non-zero absolute Doppler shift, the red shifts reported for
the core need to be adjusted up or down.

In the present paper we have made use of a more reliable method of EIS wavelength calibration developed by
\cite{young_brendan_mason}. This is based on deriving velocities using \ion{Fe}{8} in the
quiet Sun region of each slit position in the raster. We consider our method, following that of
\cite{young_brendan_mason} to be much more robust than that used by \cite{brooks_warren}. The rest of the paper is organised as follows: in section~\ref{obs} we
describe the observation used in this study; in section~\ref{analysis} we have described the data analysis and results followed in section~\ref{conc}
by a discussion of results and conclusions.

\section{Observations} \label{obs}
\begin{figure}
\centering
\includegraphics[width=0.95\textwidth]{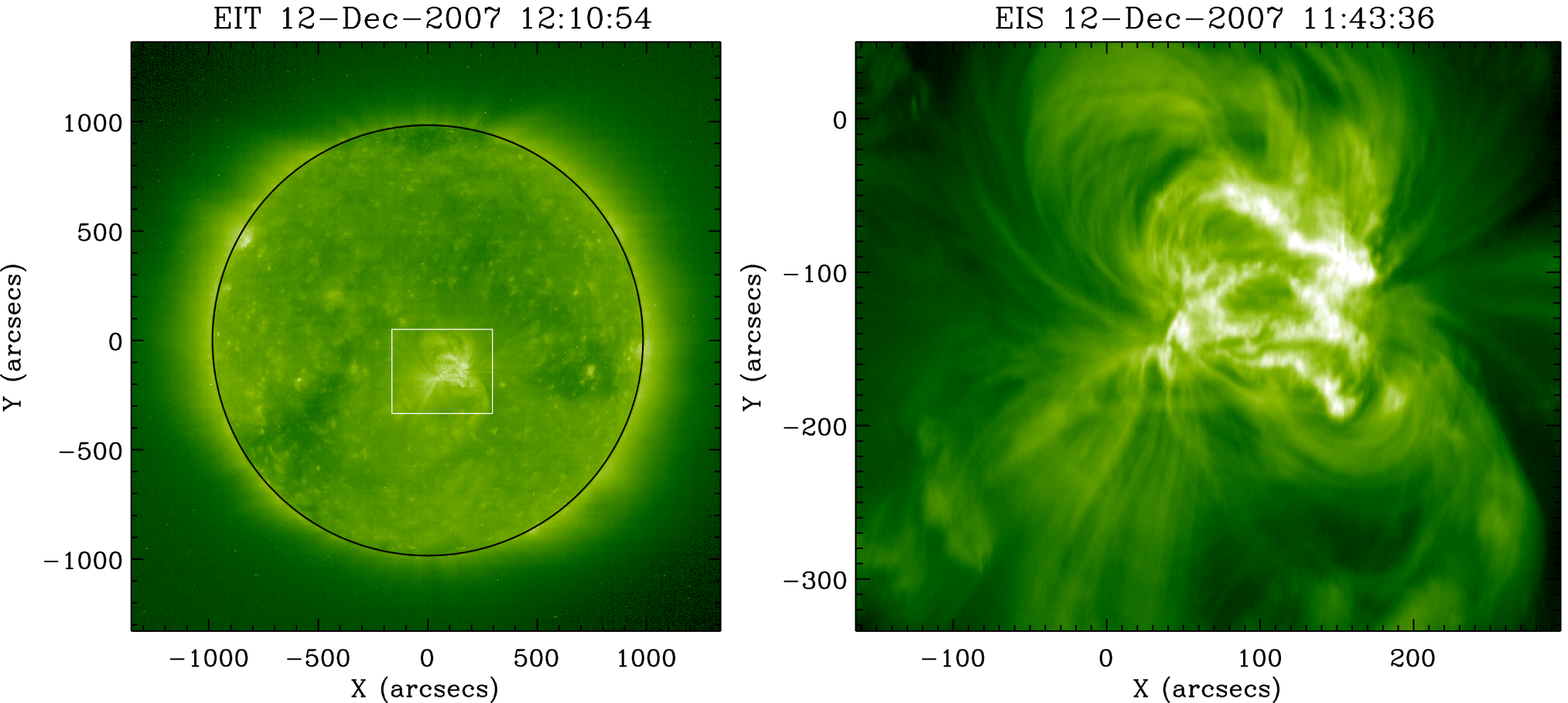}
\caption{Left panel: Full disk image of the Sun recorded by the Extreme-ultraviolet Imaging Telescope (EIT) showing the active
region located near the central meridian which was observed by the Extreme-ultraviolet Imaging Spectrometer (EIS). The over plotted box
shows the region rastered by EIS. Right Panel: A monochromatic image built in \ion{Fe}{12} line using EIS observations corresponding
to the box shown in the left panel.\label{eit_eis}}
\end{figure}

The Extreme-ultraviolet Imaging Spectrometer \citep[EIS;][]{eis}
onboard Hinode observed an active region, \textsl{AR~10978}, on
12-Dec-2007 (near the central meridian of the Sun) using the
observing sequence \textit{AR\_velocity\_map\_v2}.  This sequence
uses the 1{\arcsec}~slit with an exposure time  of 40~seconds. The
EIS raster used in this analysis started at 11:43~UT and was
completed at 17:02~UT. The left image in Fig.~\ref{eit_eis} displays
a full disk image recorded by the Extreme-ultraviolet Imaging
Telescope \citep[EIT;][]{eit} onboard the Solar and Heliospheric
Observatory (SOHO). The over-plotted box shows the area which was
rastered by EIS, for which a spectral image built in
\ion{Fe}{12}~$\lambda$195 is shown in the right panel.

\begin{figure}
\centering
\includegraphics[width=0.8\textwidth]{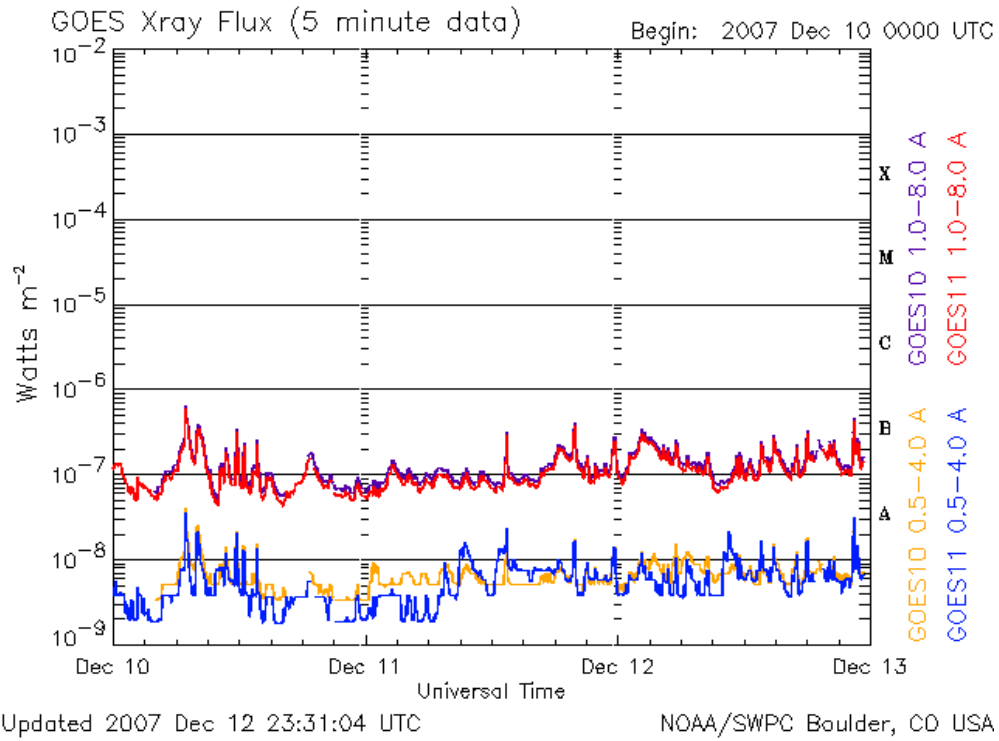}
\caption{GOES X-ray flux profile from 2007 December 10 to 13, showing minimal activity.\label{goes}}
\end{figure}

This active region raster is ideal for a study of Doppler shifts since it is very close to the central meridian and it did not show
any flaring/micro-flaring activity during the raster (as well a few days before and after the time of the observations) as shown by
the GOES plots in Fig.~\ref{goes}. The location near the central meridian is important because any small flows, which are
in the line-of-sight (LOS) can be detected with higher accuracy. It should be noted that the quiescence of the active region
ensures that our measurements are not affected by any small scale dynamic features, such as micro-flares, occurring in the core of the active region.

The EIS study sequence used for this work comprises many
different spectral lines. However, in this work we are only
concerned with the Doppler shift in the moss regions and the
temperature dependence of the Doppler shifts, so we have selected the
subset of the spectral lines from the study, given in
Table~\ref{table}. The wavelengths shown in the second column in the
table are adopted from \cite{warren_2011}. Additionally, we have
only used the spectral lines for our study which are cooler than
\ion{Fe}{13}, since the moss emission above this temperature is
highly contaminated with overlying hot core loops, as was shown by
\citet{tripathi_2008, tmdy_2010}. It should be noted that since we are only
interested in measuring the Doppler shift in the core of the active
region, we have only used the exposures covering the core of the
active region.

\begin{table}
\centering
\caption{Spectral lines used to study the the Doppler shift in the moss regions. The wavelengths shown in the second
column are taken from \cite{warren_2011}.\label{table}}
\begin{tabular}{lcc}
\hline
Ion                             &Wavelength             & log~T$_{max}$     \\
Name                        &({\AA})                    &(K)            \\
\hline
\ion{Fe}{8}             &       186.6060                & 5.60          \\
\ion{Fe}{9}             &       197.8570                & 5.85          \\
\ion{Fe}{10}            &       184.5341                    & 6.05          \\
\ion{Fe}{11}            &       188.3940                    & 6.15          \\
\ion{Fe}{12}            &       195.1186                    & 6.20          \\
\ion{Fe}{13}            &       202.0486                    & 6.25          \\
\end{tabular}
\end{table}

\section{Data Analysis and Results}\label{analysis}
We have generated intensity and Doppler maps of
the active region in the spectral lines listed in
Table~\ref{table}. As is well known, there are two instrumental
effects with EIS \citep[cf ][] {young_brendan_mason}, namely the
orbital variation of the spacecraft and the tilt of the slits with
respect to the detector. It is very important to allow for these
when determining Doppler measurements. Whilst the tilt of the slits
has been measured using independent observations and can be accounted for
uniquely from the observations, removing the orbital variation is a
non-trivial and non-unique process.

While deriving the Doppler maps, there is one additional very important matter, that of
choosing a reference wavelength. In order to measure absolute
line-of-sight velocity, we require a reference spectrum where the
Doppler shift of the line is well known. Neutral or singly ionised
photospheric or chromospheric spectral lines can be used to measure
the absolute shift, as was demonstrated by \cite{hassler} using a
rocket based spectrometer with an onboard calibration lamp. This
method is routinely used on Solar Ultraviolet Measurements of
Emitted Radiation \citep[SUMER;][]{sumer} observations \cite[see
e.g.][]{wine_2002}. Unfortunately this is not possible with the EIS
spectrum since there are no spectral lines from neutral or singly
ionized ions, with the exception of \ion{He}{2}~304 which is problematic.
\cite{young_brendan_mason} developed a method whereby they
used \ion{Fe}{8} line in the Quiet Sun region to obtain the orbital
variation and reference wavelength. We have
also used this method and the \ion{Fe}{8}~$\lambda$186.6 line
in the Quiet Sun to derive the absolute wavelength for other lines. It is then
necessary to establish the Doppler shift for the \ion{Fe}{8}~$\lambda$186.6 line in the Quiet Sun.

\cite{brooks_2011} showed that although the ionisation equilibrium calculation suggests
that the formation temperature of \ion{Fe}{8} line is $\log\,T = 5.6$, the structures seen
in \ion{Fe}{8} intensity maps show more resemblance with structures seen at  $\log\,T = 5.8$.
This fortuitously matches with \ion{Ne}{8} line for which absolute shift can be measured using SUMER observations.

Using SUMER measurements of the Quiet Sun,
\cite{pj_1999} found that \ion{Ne}{8} line formed at $\log\,T = 5.8$
are blue-shifted by 2-3~km~$s^{-1}$. On the other hand,
\cite{brekke} and \cite{chae} found that this line was red-shifted
by 5-6~km~s$^{-1}$. In a recent paper, \cite{dadashi_2011} showed
that the shift was zero at $\log\,T = 5.8$. However, we note that
the velocity given in their plot at $\log\,T = 5.8$ is not an
observational finding, but comes from three-dimensional MHD
simulations by \cite{vigo}. These measurements clearly suggest that
there are large uncertainties in absolute velocity measurements at $\log\, T = 5.8$.
Therefore, although \cite{young_brendan_mason} used the
\cite{pj_1999} values in their measurements, we have used zero shift
at $\log\,T = 5.8$, consistent with \cite{dadashi_2011}.

\begin{figure}
\centering
\includegraphics[width=0.8\textwidth]{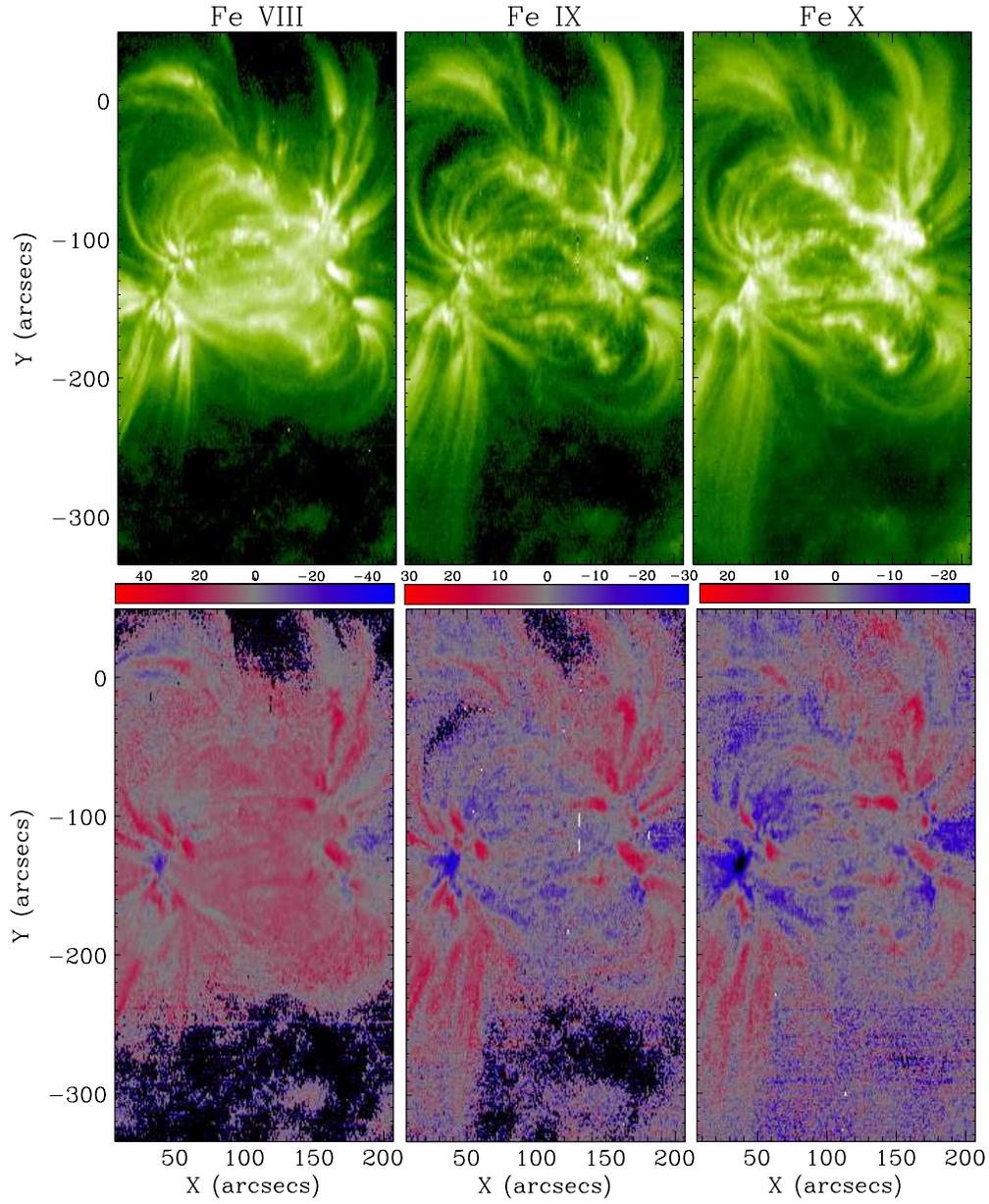}
\caption{Intensity and Doppler maps derived in \ion{Fe}{8}~$\lambda$186, \ion{Fe}{9}~$\lambda$197 and \ion{Fe}{10}~$\lambda$184.\label{Int_velocity_1}}
\end{figure}

We note here that the \ion{Fe}{8}~$\lambda$186.6 line is
blended with \ion{Ca}{14}~$\lambda$186.61 formed at $\log\,T = 6.5$
\citep{peter_2007}. While this line may have a significant blend in
the core of the active region, it has a negligible
contribution when observing in the Quiet Sun. Therefore, this line
can safely be used to derive orbital variation and rest wavelength. 
The method used can be briefly described as follows: we first
identified the lower part of the \ion{Fe}{8} raster as Quiet Sun and binned the
data in y-direction by 40 pixels (Figure~\ref{Int_velocity_1}). This increases the signal to noise ratio quite significantly. These bottom 40 pixels have
intensities equivalent to the quiet Sun intensities derived by
\cite{brooks_2009}.
We used these binned Quiet Sun data to
get orbital variation of the spacecraft and rest wavelength for
\ion{Fe}{8} line. Using the off-limb observations of
\cite{warren_2011}, and assuming that the absolute Doppler shifts
vanish for these off limb locations, we know the rest
separation of all the other lines with respect to the
\ion{Fe}{8}~$\lambda$186.6 line. This separation is used to retrieve
absolute wavelengths for all the other lines used in this study.
Using this method, \cite{young_brendan_mason} found that velocities
can be measured with an accuracy of 4$-$5 km~$s^{-1}$.

\begin{figure}
\centering
\includegraphics[width=0.8\textwidth]{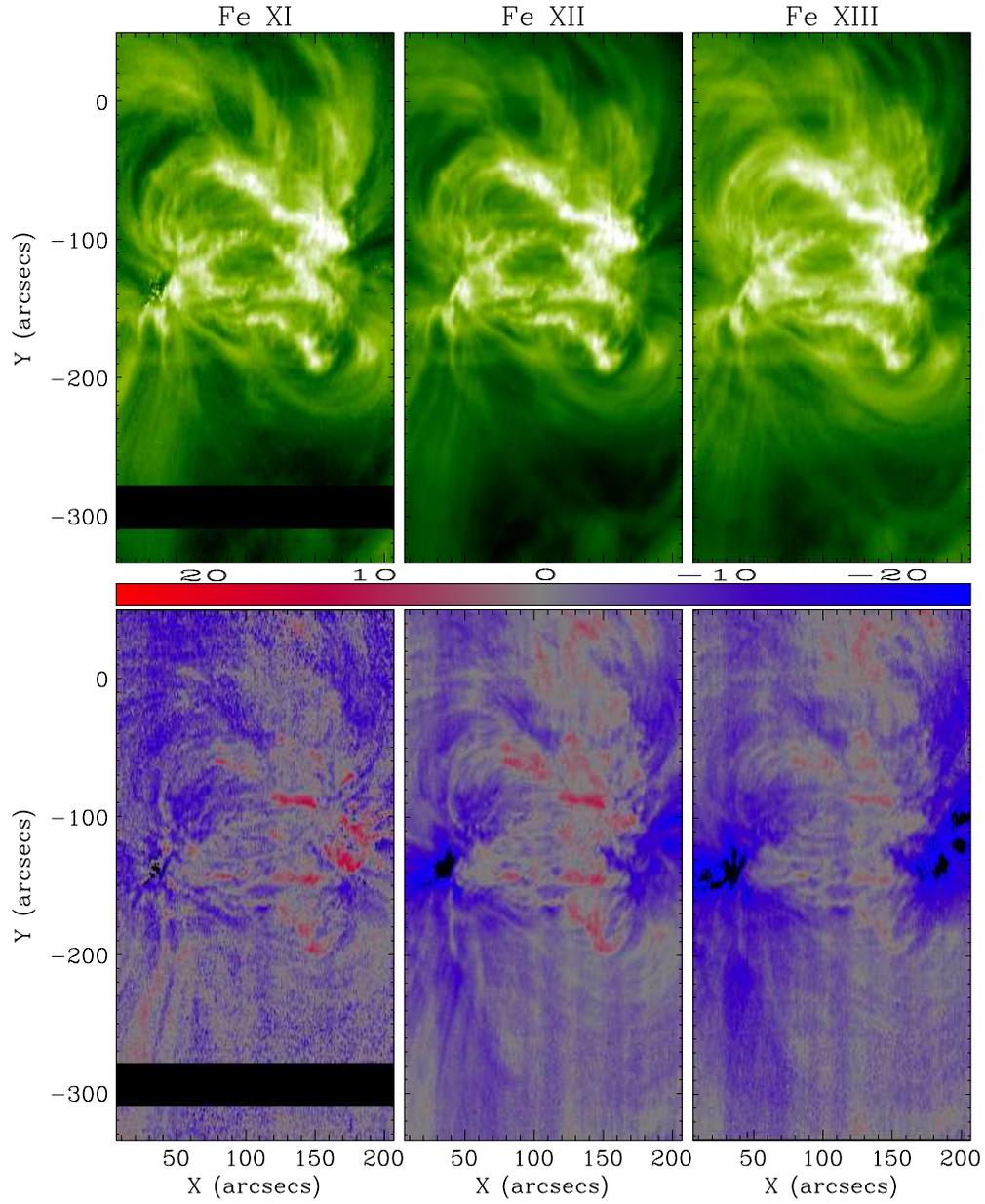}
\caption{Intensity and Doppler maps derived in \ion{Fe}{11}~$\lambda$188.3, \ion{Fe}{12}~$\lambda$195 and \ion{Fe}{13}~$\lambda$202.\label{Int_velocity_2}}
\end{figure}

Figures~\ref{Int_velocity_1} \& \ref{Int_velocity_2} display
intensity and velocity maps for the active region in the chosen
spectral lines mentioned in Table~\ref{table}. The overall intensity
structures and Doppler shift patterns are similar to those derived by previous authors, i.e., red-shifted loop structures (around 1MK) with
blue-shifted hotter regions at the boundary of the active regions
\citep[see e.g.][]{doschek_2008, delzanna_2008, marsch_2008,
tripathi_2009}. In addition, as also shown by \cite{tripathi_2009},
the footpoints of the loops are highly red-shifted at transition
region temperatures with the magnitude of the flow decreasing with
rising temperature \citep[see also][]{delzanna_2008}. As can also be
seen from the figures, the blue shifts at the boundary of the active
region increase with temperature.

As noted out earlier, \ion{Fe}{8}~$\lambda$186.6 is blended with \ion{Ca}{14}~$\lambda$186.61 which is 
formed at $\log\,T = 6.5$. The  \ion{Ca}{14}~$\lambda$186.61 line has a constant branching ratio of 0.7 
with the \ion{Ca}{14}~$\lambda$193.8. Therefore, if \ion{Ca}{14}~$\lambda$193.8 is observed in the 
spectrum, the contribution of \ion{Ca}{14}~$\lambda$186.61 can be estimated. Unfortunately, in this study 
sequence, \ion{Ca}{14}~$\lambda$193.8 was not observed. However, analysing the moss  and inter-moss 
spectra for another active region studied by \cite{tmk_2010} and \cite{tkm_2011}, we find that the 
contribution of the \ion{Ca}{14}~$\lambda$186.61 line could be as high as $20-25$\% in moss regions and 
about 50\% in the inter-moss regions. Therefore, the intensity map for \ion{Fe}{8} ~$\lambda$186.6 shown in 
the top left panel of Fig.~\ref{Int_velocity_1} may have significant contribution due to 
\ion{Ca}{14}~$\lambda$186.61 in the core of the active region. Hence, it is plausible to conclude that the 
red-shifts which are seen corresponding to the \ion{Fe}{8} line is likely to be over-estimated in the core of the 
active region. By performing a simple analytical calculation we find that if the peak of 
\ion{Ca}{14}~$\lambda$186.61 is 20\% of that of \ion{Fe}{8}~$\lambda$186.6, the overestimation in the redshift 
would be of 3.8~km~s$^{-1}$. However, if the peak of  \ion{Ca}{14}~$\lambda$186.61 is 50\% of that of 
\ion{Fe}{8}~$\lambda$186.6, then the overestimation could be as large as 6.8~km~s$^{-1}$.

The identification of the moss regions is not uniquely defined. Usually moss regions are visually
identified based on the high spatial resolution observations
recorded by TRACE in its 173~{\AA} band, which is sensitive to
\ion{Fe}{9} and \ion{Fe}{10} emission at around 1MK. Unfortunately,
there were no TRACE observations for this active region.
Although there are
EIT~171 observations essentially showing the same emission as
TRACE~173, the spatial resolution is not adequate to identify moss
regions unambiguously. Therefore, we have to rely on EIS
observations. \cite{warren_2008} identified moss regions
using an intensity threshold for the \ion{Fe}{12}~$\lambda$186.8 line
intensity, which is sensitive to electron density. We follow a similar approach.

\cite{tmdy_2010} studied the density structure in different moss regions
and found that the electron densities are greater or equal to
2$\times$10$^{9}$ cm$^{-3}$ (derived using \ion{Fe}{13} line ratio).
In this EIS study, we have two density sensitive lines of
\ion{Fe}{13}, namely $\lambda$202.04 and $\lambda$203.82, which
could be used to derive density map of the active region.

It has been found that the two CCDs of EIS are not exactly
aligned with each other \citep{peter_2007}. Since the lines we have used
in this study are all from the sort-wavelength band, this effect does not apply here.
However, due to slight misalignment between the dispersion axis of the grating and the axis
of the CCD, the images show small Y-offset \citep{peter_dens}. We have accounted for the Y-offset between
different lines using standard post processing routines provides in \textsl{SSWIDL}.

\begin{figure}
\centering
\includegraphics[width=0.95\textwidth]{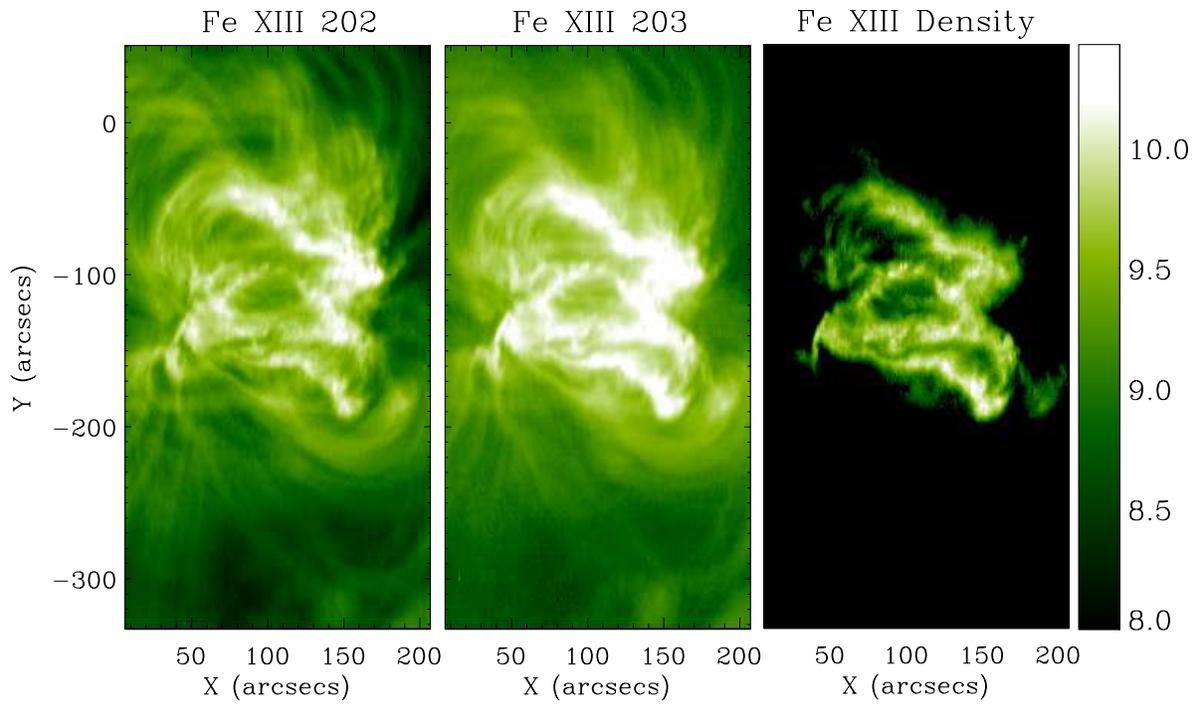}
\caption{Left panel: Intensity image obtained in \ion{Fe}{13}~202. Middle panel: Intensity image obtained in \ion{Fe}{13}~203.
Right panel: Density map using \ion{Fe}{13} line ratios.\label{fe13_dens}}
\end{figure}

\begin{figure}
\centering
\includegraphics[width=0.5\textwidth]{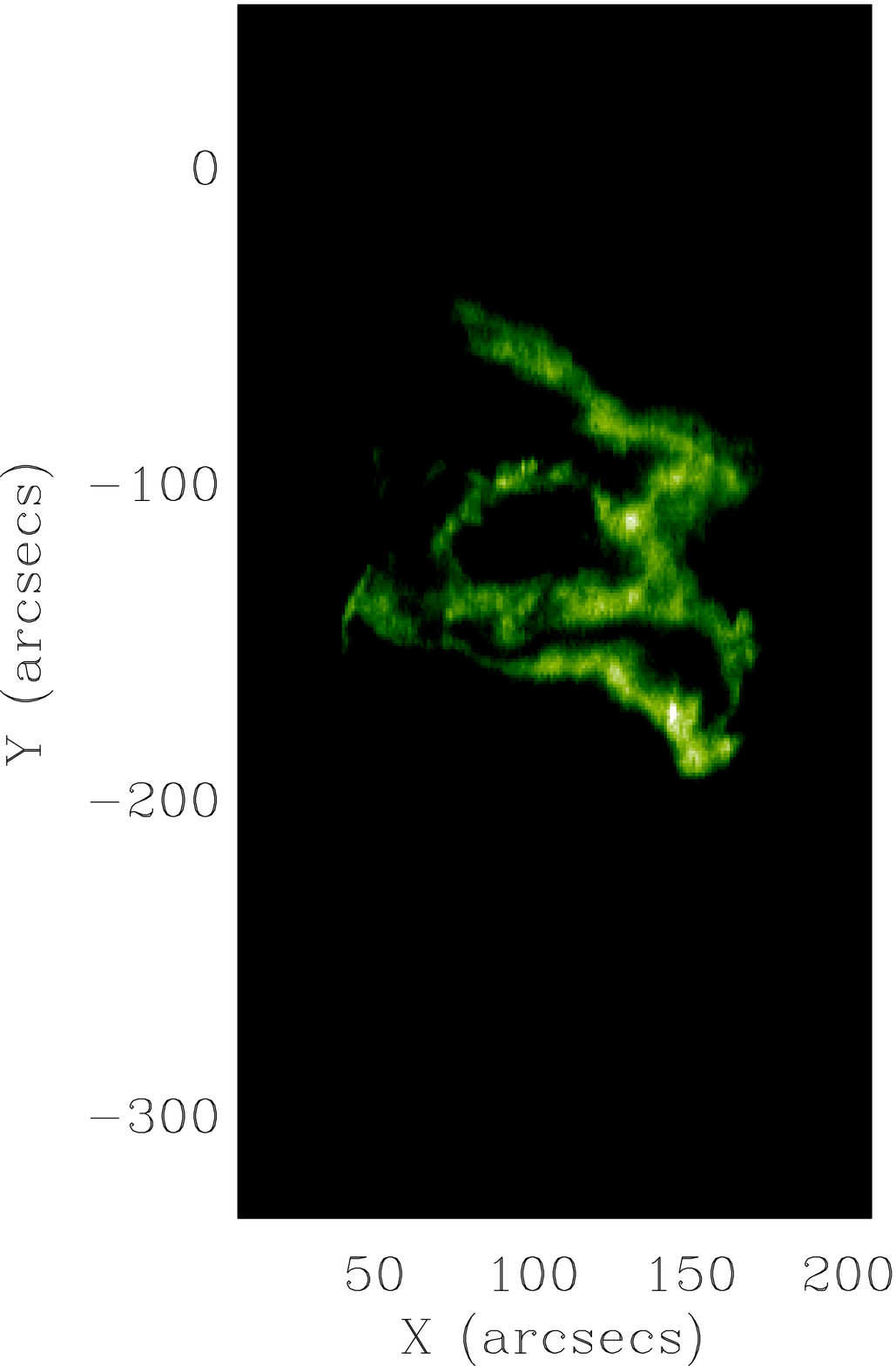}
\caption{\ion{Fe}{13} density map displayed with log~N$_e$~$\ge$~9.3~cm$^{-3}$.\label{fe13_range}}
\end{figure}

Fig.~\ref{fe13_dens} displays the intensity maps obtained in
\ion{Fe}{13}~$\lambda$202.04 (left panel) and $\lambda$203.82
(middle panel) and the corresponding density map (right panel)
derived from the ratio of these lines. The density maps are
derived using the CHIANTI v6.1 \citep{chianti_v1, chianti_v6} atomic
database. As can be seen from the intensity and density maps, the moss
regions are the densest regions in the map. However, to uniquely
obtain the locations of moss regions, we have used the electron
densities in moss regions derived by \cite{tmdy_2010} to determine a
cut-off in electron density. We have displayed the density map with log~N$_e$~$\ge$~9.3 in Fig.~\ref{fe13_range}.
We believe that these cut-off values for density uniquely
distinguish the moss region from other parts of the active region
raster.  We continue to study the Doppler shift in those pixels
only. It is important to emphasize that when extending the moss regions identified
using densities derived using \ion{Fe}{13} ratios to other lines, it is very
important to account for the Y-offset between different lines. We have
allowed for this effect in our analysis.

\begin{figure}
\centering
\includegraphics[width=0.8\textwidth]{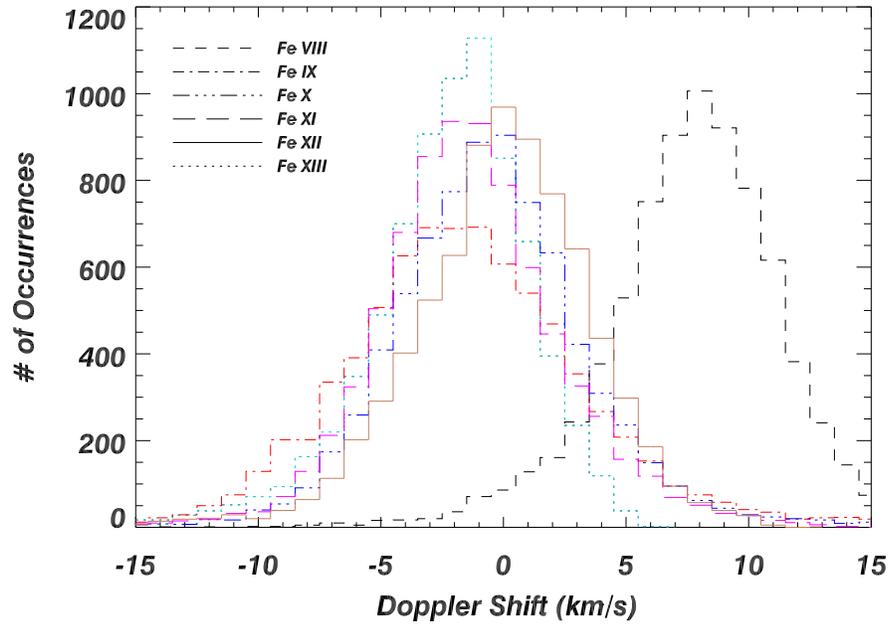}
\caption{Histogram of the velocities corresponding to the pixels which have densities log~N$_e$~$\ge$~9.3~cm$^{-3}$.\label{velocity_hist}}
\end{figure}

Figure~\ref{velocity_hist} displays the histograms of the Doppler
shift for the pixels with densities with log~N$_e$~$\ge$9.3 (shown
in Fig.~\ref{fe13_range}) for all the spectral lines listed in
Table~\ref{table}. As can be very clearly seen, all the histogram
plots peak very close to zero on the negative velocity (blue) side,
with the exception of the \ion{Fe}{8} line (which is blended with
the \ion{Ca}{14} line). We note here, however, that if we assume 
that contribution of \ion{Ca}{14} is as high as 20\% and subtract 
4~km~s$^{-1}$ from the \ion{Fe}{8} velocity histogram, the histogram 
will peak at around 5~km~s$^{-1}$. This is similar to what is found in 
\ion{Ne}{8}, which forms almost at the same temperature to that of \ion{Fe}{8}, 
observed in Moss regions using SUMER spectra 
(Winebarger et al. in preparation).The velocity distribution for all other
lines peaks close to zero. With increasing temperature, the
histogram starts to become asymmetric with more points on the blue
side. Table~\ref{table:2} provides the mean, median and standard
deviation values of the measured velocity distributions for all the
lines in the moss region. The absolute mean velocities are negative
i.e. blue shift, which is increasing with temperature. We have also
computed the second moment of the velocity distribution i.e.
standard deviation given in the 5th column of the table. The
standard deviation decreases with temperature. Since the mean and
median of the measured velocity distributions were within the
systematic errors on the measurements, we wanted to find out how
many moss pixels have definite shifts. For this purpose we computed
the fraction of pixels with velocity amplitudes
greater than the 5~km~s$^{-1}$ uncertainty (shown in the 2nd to last
column of the Table~\ref{table:2}).  We can also take the highly
conservative view that the absolute velocity calibration is unknown.
In that case, the fraction of pixels that deviate from the mean by
more than the 5~km~s$^{-1}$ uncertainty (last column in the table)
is the smallest possible fraction of pixels that must have a
non-zero Doppler shift. The subscript 'i' represents the moss
pixels.  Both of these measures indicate that a sizable fraction of
moss pixels ($\approx 30\%$ for \ion{Fe}{9}) cannot be at rest.

It is worthwhile mentioning that there may be a few pixels qualifying as moss (based on the density cut-off definition) affected by the tilted point spread function of EIS \citep[see e.g..][]{young_brendan_mason, sm_2012, tmd_2012}.
Due to this effect, the intensity and corresponding velocity signatures
may be shifted by 1~pixel when there is a sharp change in intensity. Therefore, although the pixel in the intensity map fulfils the criteria of
moss, the velocity signature of that pixel would be moved to
another adjacent pixel which may not fulfil the criteria.

\begin{table}
\centering
\caption{Mean, Median and Standard Deviation of the Doppler shift measured in moss regions and fraction of mossy pixels with definite Doppler shift.\label{table:2}}
\begin{tabular}{l|c|ccc|c|c}
\hline
Ion Name                                & log~T$_{max}$                     &\multicolumn{3}{|c|}{Doppler shift in moss}                      & $|v_i|>5$           &      $|v_i-<v>|>5$    \\
                                                                                            \cline{3-5}                                                                                                         \\
                                                &   (K)                                     & Mean                  &       Median              & Standard      &                                       \\
                                            &                                   &   $<v>$               &                               & Deviation      &                                      \\
                                                &                                           & km~s$^{-1}$       & km~s$^{-1}$               & km~s$^{-1}$ & (\%)            &      (\%) \\
\hline
\ion{Fe}{9}                             & 5.85                                      &$-$0.6                 &   $-$1.7                      &4.9          & 31.0            &               28.7\\
\ion{Fe}{10}                            & 6.05                                      &$-$0.6                 &   $-$0.6                      &4.0          & 18.7            &               17.9\\
\ion{Fe}{11}                            & 6.15                                      &$-$1.4                 &   $-$1.5                      &3.8          & 19.5            &               16.9\\
\ion{Fe}{12}                            & 6.20                                      &$-$0.1                 &   0.9                             &3.7          & 16.0            &               16.0\\
\ion{Fe}{13}                            & 6.25                                      &$-$2.3                 &   $-$1.9                      &3.6          & 17.7            &               12.0\\
\hline
\end{tabular}
\end{table}

\section{Summary and Discussion} \label{conc}

We have studied the Doppler shift and temperature dependence of the
Doppler shift in moss regions using an active region observation
recorded on 12-Dec-2007. We have distinguished the moss regions from
the rest of the active region by defining a low density cut-off as
was derived by \cite{tmdy_2010}.  The histograms of the velocity
distributions peak at around 0~km~s$^{-1}$. This is true for all the
spectral lines studied in this work, except the \ion{Fe}{8} line
which is blended with a \ion{Ca}{14} line in its red-wing leading to
unreliable results in the active region core. The mean of the
distributions show blue shifts which increase with temperature. With
increasing temperature, the width of the distribution decreases and
the distribution becomes asymmetric towards the blue. In
addition, there are significant number of pixels showing velocities
larger than the error i.e. 5~km~s$^{-1}$ shown the last two columns
of Table~\ref{table:2}.

Using analytical modeling, \cite{steve_2008} suggested that
the radiation from moss regions is powered by an enthalpy flux from
cooling and draining of coronal plasma. He predicted that downflow
speeds at $T = 1MK$ ranging between 2.9 and 3.4 km s$^{-1}$ should be observed. We observe Doppler shifts close to zero. However, we
note that we have a large uncertainty of about 4-5~km~s$^{-1}$
in our measurements. The observed line profiles are a composite
made up of red-shifted
emission from a cooling downflow and faster blue-shifted emission
from either evaporating plasma associated with nanoflares
\citep{pk_2006} or hot plasma at the tips of Type-II spicules
\citep{depontieu_2011}. The analysis of \cite{steve_2008} applies
only to the dominant red-shift component. As the blue-shift
component becomes progressively brighter, as predicted for higher
temperatures \citep{pk_2006}, it will displace a Gaussian fit of the
line profile further to the blue, consistent with our findings. Further work is in progress relating to this effect, using models with
nonequilibrium ionization \citep{bk_2011}.

There have been some previous measurements of Doppler shifts
in moss regions.  \cite{klimchuk_1987} found that active region
plage (essentially an early term for moss) are red shifted by
approximately 10-15~km~s$^{-1}$ in the \ion{C}{4}~$\lambda$1548 line
formed at 0.1 MK.  This is consistent with the picture that red
shifts are strongest at low temperatures and become progressively
weaker at higher temperatures, eventually transitioning to blue
shifts. Our observations appear to bracket this range of transition.
The trend from red shift to blue shift with increasing temperature has
been reported previously for the quiet Sun \citep{pj_1999}, and it
is now clear that it also applies to active region moss.  However, the
transition from red to blue appears to occur at a higher temperature
in moss ($\approx$1 MK versus 0.5 MK in the quiet Sun). Our preliminary
modeling results suggest that this can be explained by higher energy
nanoflares in active regions (Bradshaw and Klimchuk, in
preparation).

Although our results are consistent with impulsive heating model, we
cannot rule-out steady state heating scenario, where conductive flux
from the corona is balanced by radiative cooling. This basically
suggests no bulk flow i.e. 0~km~$s^{-1}$ \citep[although see
e.g.,][]{mariska_boris}, which is consistent with what we observe in
this paper for the peak of the velocity distribution. However, we
also observe apparent asymmetries in the distribution with
increasing temperature, which is not predicted from the steady
heating scenario. It is worthwhile noting here that our
uncertainties in the absolute velocities are so large
(4$-$5~km~s$^{-1}$) that we cannot rule-out one model against other
for most pixels, solely based on the Doppler shift measurements
using EIS data. However, the result that there are significant
number of pixels with velocity amplitudes larger than 5~km~s$^{-1}$
indicates that the heating is not steady at those locations and may
suggest that impulsive heating is more likely in general.

Further, more accurate measurements of Doppler shifts in moss regions are urgently required to distinguish between the different heating scenarios. We note
that NASA's Interface Region Imaging Spectrometer (IRIS) is to be launched in late 2012 and should provide some conclusive results, in particular if IRIS
observations are combined with EIS studies of moss regions. In addition, ESA's Solar Orbiter is to be launched in 2017 and the planned Solar-C mission  will
provide unprecedented opportunities to perform such Doppler shift measurements with higher accuracy.

\acknowledgments{We thank an anonymous referee for constructive comments which has improved the paper.We acknowledge useful discussions at the ISSI on Active Region Heating. We also acknowledge the
loops workshops as an opportunity to stimulate discussions and
collaborate on this project.
HEM acknowledges support from STFC and JAK
acknowledges support from the NASA Supporting Research and
Technology Program. We thank Dr Peter Young for various discussions. Hinode is a Japanese mission developed and
launched by ISAS/JAXA, collaborating with NAOJ as a domestic
partner, NASA and STFC (UK) as international partners. Scientific
operation of the Hinode mission is conducted by the Hinode science
team organised at ISAS/JAXA. This team mainly consists of scientists
from institutes in the partner countries. Support for the
post-launch operation is provided by JAXA and NAOJ (Japan), STFC
(U.K.), NASA, ESA, and NSC (Norway). CHIANTI is a collaborative
project involving researchers at NRL (USA) RAL (UK), and the
Universities of: Cambridge (UK), George Mason (USA), and Florence
(Italy).}



\end{document}